\documentclass[10pt,preprint,showpacs,showkeys,preprintnumbers,superscriptaddress,amsmath,amssymb,twocolumn,prb]{revtex4}
\usepackage{graphicx}
\usepackage{dcolumn}
\usepackage{bm}
\usepackage{epsfig}
\usepackage{amsmath}
\usepackage{amsfonts}
\usepackage{amssymb}
\usepackage{color}

\begin{document}

\preprint{}
\title{Mg/Ti multilayers: structural, optical and hydrogen absorption properties}
\author{A.~Baldi}
\affiliation{Department of Physics and Astronomy, VU University Amsterdam, De Boelelaan 1081, 1081 HV Amsterdam, The Netherlands}
\email[Corresponding author email: abaldi@few.vu.nl]{}
\author{G.~K.~P\'{a}lsson}
\affiliation{Department of Physics, Uppsala University, Box 530, S-751 21 Uppsala, Sweden}
\author{M.~Gonzalez-Silveira}
\author{H.~Schreuders}
\author{M.~Slaman}
\author{J.~H.~Rector}
\affiliation{Department of Physics and Astronomy, VU University Amsterdam, De Boelelaan 1081, 1081 HV Amsterdam, The Netherlands}
\author{G.~Krishnan}
\author{B.~J.~Kooi}
\affiliation{Zernike Institute for Advanced Materials, University of Groningen, Nijenborgh 4, 9747 AG Groningen, The Netherlands}
\author{G.~S.~Walker}
\affiliation{Faculty of Engineering, University of Nottingham, University Park, Nottingham NG7 2RD, UK}
\author{M.~W.~Fay}
\affiliation{Nottingham Nanotechnology and Nanoscience Centre, University of Nottingham, Nottingham NG7 2RD, UK}
\author{B.~Hj\"{o}rvarsson}
\affiliation{Department of Physics, Uppsala University, Box 530, S-751 21 Uppsala, Sweden}
\author{R.~J.~Wijngaarden}
\affiliation{Department of Physics and Astronomy, VU University Amsterdam, De Boelelaan 1081, 1081 HV Amsterdam, The Netherlands}
\author{B.~Dam}
\affiliation{DelftChemTech, Delft University of Technology, Julianalaan 136, 2600 GA Delft, The Netherlands}
\author{R.~Griessen}
\affiliation{Department of Physics and Astronomy, VU University Amsterdam, De Boelelaan 1081, 1081 HV Amsterdam, The Netherlands}

\date{\today}
\preprint{}
\begin{abstract}
Mg-Ti alloys have uncommon optical and hydrogen absorbing properties, originating from a ``spinodal-like'' microstructure with a small degree of chemical short-range order in the atoms distribution. In the present study we artificially engineer short-range order by depositing Pd-capped Mg/Ti  multilayers with different periodicities and characterize them both structurally and optically. Notwithstanding the large lattice parameter mismatch between Mg and Ti, the as-deposited metallic multilayers show good structural coherence. Upon exposure to H$_2$ gas a two-step hydrogenation process occurs, with the Ti layers forming the hydride before Mg. From \textit{in-situ} measurements of the bilayer thickness $\Lambda$ at different hydrogen pressures, we observe large out-of-plane expansions of the Mg and Ti layers upon hydrogenation, indicating strong plastic deformations in the films and a consequent shortening of the coherence length. Upon unloading at room temperature in air, hydrogen atoms remain trapped in the Ti layers due to kinetic constraints. Such loading/unloading sequence can be explained in terms of the different thermodynamic properties of hydrogen in Mg and Ti, as shown by diffusion calculations on a model multilayered systems. Absorption isotherms measured by hydrogenography can be interpreted as a result of the elastic clamping arising from strongly bonded Mg/Pd and broken Mg/Ti interfaces.
\end{abstract}



\maketitle

\section{Introduction}

Magnesium and titanium are immiscible. Metastable Mg-Ti alloys have nevertheless been successfully prepared in thin films by high energy processes, such as electron beam deposition \cite{Niessen2005} and magnetron sputtering.\cite{BorsaAPL2006,baoSEMSC2008} These films have gravimetric hydrogen storage capacities up to 6.5 wt\%{} \cite{Niessen2005} and fast and reversible kinetics of hydrogen absorption and desorption.\cite{BorsaAPL2006} The structural reversibility of Mg-Ti alloys is particularly surprising when considering the strong segregation occurring in many Mg-based binary systems.\cite{GiebelsPRB2004,Gonzalez-Silveira2008}
Furthermore, when exposed to H$_2$, Pd-capped Mg-Ti thin films switch from a reflecting metallic state to a black, light-absorbing, hydrogenated state.\cite{BorsaAPL2006} This reversible optical black state can be applied in hydrogen sensors \cite{SlamanSAB2007} and smart absorbers for solar collectors.\cite{BorsaAPL2006,baldiIJHE2008} As also suggested by first principle calculations,\cite{vanSettenPRB2009} the black appearance of the hydride is due to the formation of a face-centered-cubic phase, in which Mg and Ti atoms are distributed among the lattice sites with a certain degree of chemical short-range order,\cite{gremaudPRB2008,baldiIJHE2009} leading to the coexistence of Mg-rich and Ti-rich regions with structurally coherent boundaries. Such ``spinodal-like'' microstructure is not uncommon in immiscible binary alloys \cite{Ma2005,hePRL2001} and it is a key ingredient in understanding the exceptional reversibility of Mg-Ti thin films.\cite{BorsaPRB2007,baldiIJHE2009} In order to achieve a deeper understanding of the role of chemical segregation on the structural, optical and hydrogen absorbing properties of these systems, we engineered one dimensional short-range order by depositing several Mg/Ti multilayers with different periodicities. By means of optical and structural studies we are able to reconstruct the hydrogen loading sequence, measure the out-of-plane expansion of the individual layers and detect the breaking of structural coherence occurring at the Mg/Ti interfaces upon formation of TiH$_2$. Such removal of coherence is responsible for the ``scissor'' effect observed in Ti-sandwiched Mg thin films:\cite{baldiAPL2009} when a thin Mg film is sandwiched between Ti layers it absorbs hydrogen at pressures close to bulk Mg, effectively behaving as quasifree. This is due to the fact that Mg and Ti, thanks to their positive enthalpy of mixing, form interfaces with poor adhesion, which become even more disconnected when Ti expands upon hydrogen absorption. On the contrary, in Mg films capped with Pd, alloying occurs at the Mg/Pd interface and Mg feels an elastic constraint due to the presence of the cap layer that leads to higher equilibrium pressures of hydrogen absorption.\cite{baldiPRL2009}

\section{Experimental details}

\begin{figure}[htbp]
\begin{center}
\includegraphics[width=8.6 cm]{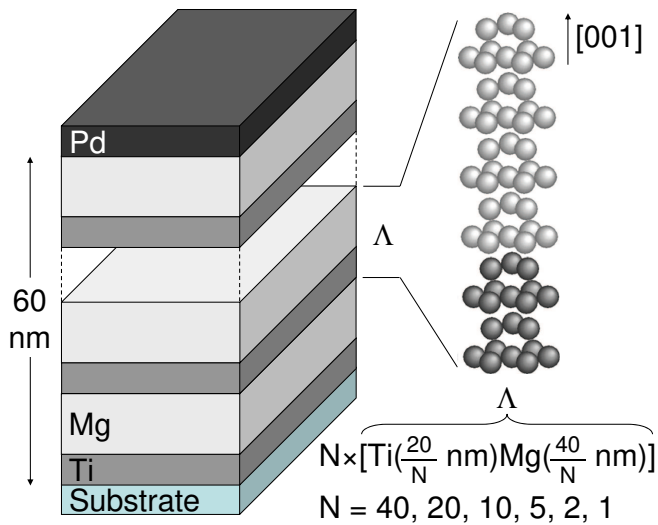}
\caption{(color online) Sample geometry. The total multilayers thickness is 60 nm and, in every sample, Mg layers are twice as thick as Ti ones: Nx[Ti($\frac{20}{\mathrm{N}}$ nm)Mg($\frac{40}{\mathrm{N}}$ nm)], with N = 40, 20, 10, 5, 2 and 1.}\label{cartoonmultilayers}
\end{center}
\end{figure}

Mg/Ti multilayers covered with Pd are deposited in a ultra-high vacuum (UHV) compatible system (base pressure = $10^{-6}$ Pa) equipped with a computer controlled shutter system, by DC and RF magnetron sputtering of Mg (99.95\%{}), Ti (99.999\%{}) and Pd (99.98\%{}) targets in 0.3 Pa of Ar, on substrates kept at room temperature. The substrates used are 10x10x1 mm polished single-crystal Si(100) for XRD, XRR and HRTEM, silicon nitride membranes for in-plane TEM, 20x10x1 mm quartz for optical spectroscopy and 10x10x0.5 mm float glass for hydrogenography measurements. In order to obtain homogenous films the substrates are continuously rotated during sputtering. The films are covered with Pd to prevent oxidation and promote hydrogen dissociation and absorption. The Pd thickness varies between 1 and 10 nm depending on the experimental technique used to analyze the samples. Typical deposition rates are $0.22~\mathrm{nm/s}$ for Mg at 150~W (RF), 0.08 $\mathrm{nm/s}$ for Ti at 200~W (DC) and $0.11~\mathrm{nm/s}$ for Pd at 50~W (DC). In each Mg/Ti sample the Mg layers are twice as thick as the Ti ones, giving, after correcting for the molar volumes ($\overline V_{\mathrm{Mg}}=13.97$ cm$^3$/mol, $\overline V_{\mathrm{Ti}}=10.64$ cm$^3$/mol), an overall composition of Mg$_{0.60}$Ti$_{0.40}$. The total multilayer thickness is 60 nm with 6 different bi-layer thicknesses $\Lambda$, ranging from 1.5 to 60 nm: Nx[Ti($\frac{20}{\mathrm{N}}$ nm)Mg($\frac{40}{\mathrm{N}}$ nm)], with N = 40, 20, 10, 5, 2 and 1. These samples characteristics allow us to have fast and comparable kinetics of hydrogen absorption and desorption in all the multilayers. A sketch of the samples geometry is shown in Fig. \ref{cartoonmultilayers}.

XRD patterns are measured in a $\theta-2\theta$ configuration, with a Bruker D8 Discover diffractometer equipped with a two-dimensional detector for real-time data collection over a large area with high sensitivity and low background. A beryllium dome allows \textit{in-situ} diffraction measurements during hydrogenation/deydrogenation of the films in hydrogen pressures up to 10$^5$ Pa and temperatures between room temperature and 473 K.

XRR measurements are performed on the 10x sample in a specially designed UHV chamber mounted on a Bruker Discover D8 X-ray diffractometer equipped with a parallel X-ray beam (CuK$\alpha_1$ $\lambda=0.15406$ nm), that allows \textit{in-situ} hydrogen loading in a wide range of temperatures and pressures.\cite{Palsson2008} The sample is measured at 333 K both in the as-deposited metallic state and during hydrogen uptake at different H$_2$ pressures. The temperature is high enough to promote fast kinetics of hydrogen absorption but low enough to avoid severe alloying at the Mg/Pd interface. In-plane resistance measurements are used to determine whether the dissolved hydrogen is in equilibrium with the surrounding H$_2$ atmosphere.

HRTEM was performed on a JEOL 2100F FEG-TEM, equipped with Digital STEM, Gatan Orius imaging system, and Gatan DigiScan. Cross section samples were prepared by ion beam milling using an FEI Quanta 200 ED FIB-SEM and transferred to the TEM using a Gatan HHST 4004 environmental cell and vacuum transfer holder to minimise oxidation.

Hydrogen loading isotherms are measured at 333 K by means of Hydrogenography,\cite{Gremaud2007ADM} an optical technique that allows to detect the amount of light transmitted through a thin film, while slowly increasing the hydrogen pressure at constant temperature. The  Pressure-Optical transmission-Isotherms (PTIs) obtained by hydrogenography can be directly related to the standard Pressure-Composition-Isotherms (PCIs) measured for metal hydrides.\cite{Gremaud2007APL} Details of the hydrogenography experimental setup can be found in Gremaud \textit{et al.}.\cite{Gremaud2007ADM}

Optical spectra are measured with a Perkin Elmer Lambda 900 diffraction grating spectrometer ($0.5<\hbar\omega<6.5$ eV). Reflection and transmission spectra of the as-deposited and hydrogenated films are measured through the transparent quartz substrate at 333 K in vacuum and in 10$^5$ Pa H$_2$, respectively.

\section{Results and discussion}

\subsection{Structural characterization: XRD, XRR and TEM}

\subsubsection{As-deposited multilayers}

\begin{figure}[htbp]
\begin{center}
\includegraphics[width=8.6 cm]{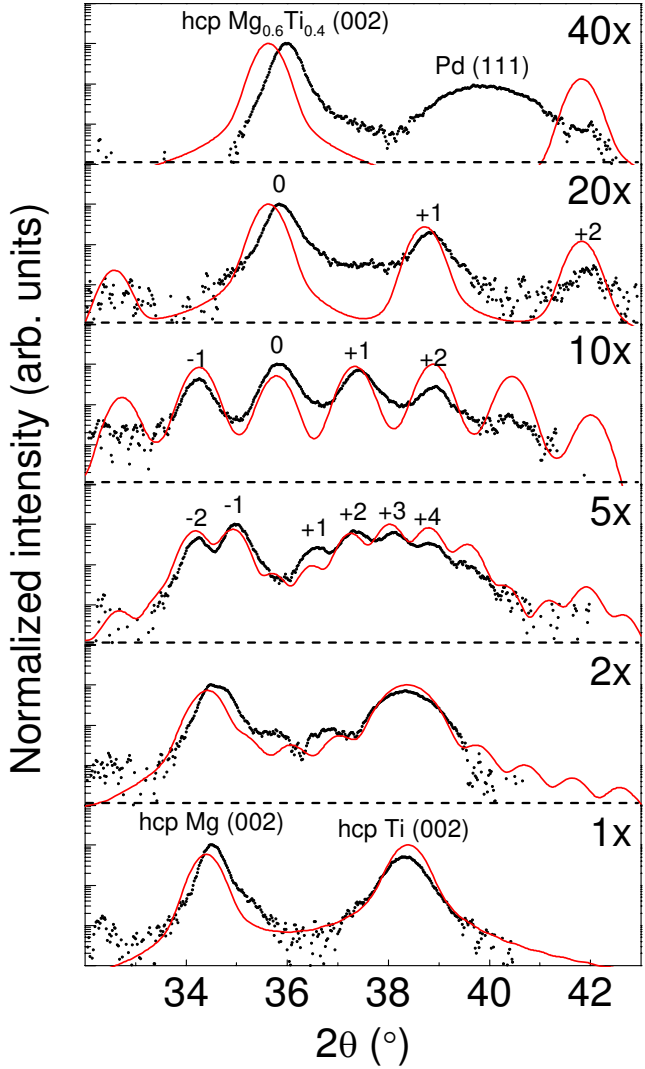}
\caption{(color online) Measured (dots) and simulated (lines) XRD patterns of Nx[Ti($\frac{20}{\mathrm{N}}$ nm)Mg($\frac{40}{\mathrm{N}}$ nm)] multilayers with N = 40, 20, 10, 5, 2 and 1, covered with 5 nm of Pd. The simulations are based on idealized step model, not capturing the influence of defects and other imperfections.}\label{XRDlog}
\end{center}
\end{figure}

The uncorrected diffraction patterns measured for the as-deposited samples in vacuum at room temperature are shown in Fig. \ref{XRDlog}.
Both Mg and Ti have hexagonal-closed-packed structures in their elemental form and, as will be shown below, the Mg/Ti multilayers grow with the (001) plane parallel to the substrate surface. For the 1x sample the Ti and Mg layers are thick enough to give the reflections of pure elements, while already for the 5x sample superlattice peaks (satellite peaks) begin to appear, due to the repetitions of the bi-layer thickness $\Lambda$. The broad peak appearing at $\sim$40$^{\circ}$ for the 40x sample is the (111) reflection from the face-centered cubic Pd. The absence of any Pd reflection
in the patterns of the other multilayers is not surprising when considering that Pd is deposited on top of the uppermost Mg layer. Pd deposited on Mg$_y$Ti$_{1-y}$ thin films, in fact, does not produce any diffraction signal for $y>0.9$, although its presence is confirmed by Rutherford Backscattering Spectrometry (RBS).\cite{BorsaPRB2007} This peak disappearance is most likely due to the increase in
lattice mismatch between Pd and Mg$_y$Ti$_{1-y}$ with increasing magnesium content, suggesting that for the 40x sample the Ti and Mg layers
are thin enough to undergo lattice deformations similar to the ones expected in a Mg$_{0.6}$Ti$_{0.4}$ alloy. In Fig. \ref{XRDlog} we also show the simulated patterns obtained with an ideal step model, which assumes perfect superlattices with a square-wave composition modulation along the growth direction, and coherent Mg/Ti interfaces.\cite{jin1989ADVPHYS,MichaelsenPMA1995} In the model the diffracted intensity $I$ is given by:
\begin{equation}\label{michaelsen}
I=I_{\mathrm{N}}\left(I_{\mathrm{Mg}}+I_{\mathrm{Ti}}+I_{\mathrm{MgTi}}\right)
\end{equation}
where $I_{\mathrm{N}}$ is a term due to the N bilayer repetitions in the multilayers, $I_{\mathrm{Mg}}$ and $I_{\mathrm{Ti}}$ are the intensities of the consituent materials and $I_{\mathrm{MgTi}}$ is a mixed term arising from the structural coherence.\cite{MichaelsenPMA1995} In incoherent multilayers the mixed term disappears leading to a different distribution of intensities in the satellite peaks. The scattering powers of the elements are approximated with their atomic numbers and the patterns are filtered with a Gaussian distribution, with a full width at half maximum (fwhm) of $0.47^{\circ}$, to take into account the instrumental broadening and the possible deviations from a perfect geometry. These deviations include random variations in the number of atomic layers, interface roughness or interdiffusion, variations in the lattice spacings due to in-plane elastic coherency strain, distribution of sizes in in-plane grains and random orientations of the growth direction of each grain. Several models have been developed to include corrections to the ideal step model and take into account deviations from a perfect geometry.\cite{jin1989ADVPHYS,fullertonPRB1992} In our XRD measurements, however, the experimental broadening is very high, of the order of $0.3^{\circ}$, due to the use of a two-dimensional detector that has a finite grid and requires a rather large beam spot on the sample in order to produce a high signal-to-noise ratio. Such a large experimental broadening hinders any attempt to obtain more detailed informations from the XRD patterns. Nevertheless, the simulations shown in Fig. \ref{XRDlog} qualitatively reproduce all the features observed experimentally, showing that upon deposition Mg and Ti form well defined layers with partially coherent interfaces. The structural coherence in Mg/Ti multilayers is rather surprising given the 8.7\%{} lattice mismatch between the in-plane cell parameter of Mg (a$_{\mathrm{hcp}}=0.3209$ nm) and Ti (a$_{\mathrm{hcp}}=0.2951$ nm). The rocking curve measured over the fundamental peak ($s=0$) of the 10x sample has a fwhm of 5.7$^{\circ}$, indicating a moderately textured multilayer.

\begin{figure}[htbp]
\begin{center}
\includegraphics[width=8.6 cm]{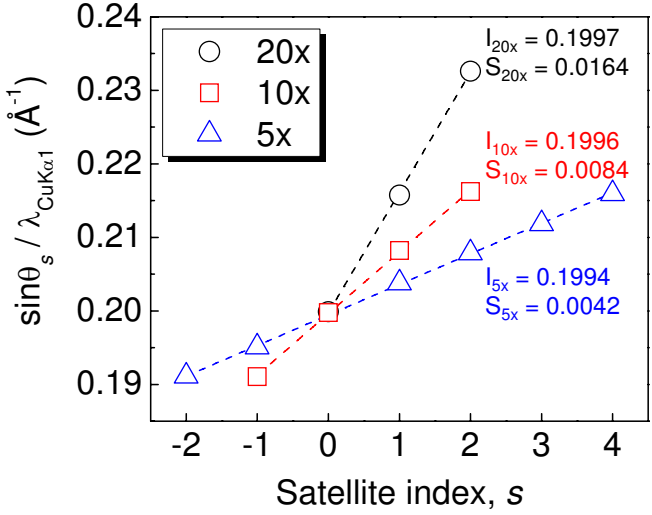}
\caption{(color online) Measured linear dependence of the ratio $\sin\theta_s/\lambda_{\mathrm{CuK\alpha_1}}$ on the satellite index $s$ for the 20x (circles), 10x (squares) and 5x (triangles) samples as indicated in Fig. \ref{XRDlog}. The dashed lines are linear fits to the data with slopes S and intercepts I.}\label{lambdafit}
\end{center}
\end{figure}

For a multilayer the Bragg law can be written as:\cite{Palsson2008}
\begin{equation}\label{Bragg}
\frac{\sin\theta_s}{\lambda_{\mathrm{CuK\alpha_1}}}=\frac{1}{2d_{\mathrm{Mg/Ti}}}\pm\frac{s}{2\Lambda}
\end{equation}
where $s$ is the index of the satellite peak with respect to the fundamental peak ($s=0$), as shown for samples 20x, 10x and 5x in Fig. \ref{XRDlog}, $d_{\mathrm{Mg/Ti}}$ is the average interplanar distance of a Mg/Ti bilayer and $\Lambda$ is the bilayer thickness.
In Fig. \ref{lambdafit} we plot the ratio $\sin\theta_s/\lambda_{\mathrm{CuK\alpha_1}}$ versus $s$, for the satellite peaks of the 20x, 10x and 5x samples.
Using Eq. \ref{Bragg}, from the fitted slopes, S, and intercepts, I,  we obtain the bilayer thicknesses $\Lambda$ and the average interplanar distance $d_{\mathrm{Mg/Ti}}$, see Table \ref{lambdafittable}.
\begin{table}[htbp]
\caption{Comparison between the experimental and nominal values for the bilayer thickness $\Lambda$ and the average interplanar distance $d_{\mathrm{Mg/Ti}}$ obtained from eq. \ref{Bragg} for samples 20x, 10x and 5x.}\label{lambdafittable}
\begin{tabular}{c|c|c|c|c}
Sample& $\Lambda(\mathrm{nom})$ & $\Lambda(\mathrm{exp})$ & $d_{\mathrm{Mg/Ti}}(\mathrm{nom})$ & $d_{\mathrm{Mg/Ti}}(\mathrm{exp})$\\
&nm&nm&nm&nm\\
\hline
20x & 3.0 & 3.0 & 0.2518& 0.2504\\
10x & 6.0 & 6.0 & 0.2518& 0.2505\\
5x & 12.0 & 11.9 & 0.2518& 0.2508\\
\end{tabular}
\end{table}
The excellent agreement between experiment and model indicates that the textured regions of our Mg/Ti multilayers correspond closely to their ideal geometries.

A rough estimate of the average coherence length $\xi$, defined as the distance over which the atomic positions are quantitatively correlated,\cite{fullertonPRB1992} can be obtained with Scherrer's formula: $\xi=(K\cdot\lambda_{\mathrm{CuK\alpha_1}})/(\beta\cdot\cos\theta)$, where $K$ is the shape factor ($K$ ranging between 0.9 and 1, depending on the grains shape) and $\beta$ is the fwhm expressed in radians. For the fundamental peak ($s=0$) of the 10x sample we have $\beta\approx0.5^{\circ}=8.7\cdot10^{-3}\ \mathrm{rad}$ and $\theta=17.92^{\circ}$, corresponding to a coherence length $\xi \approx17$ nm, assuming spherical grains. Given the very high experimental broadening in our XRD measurements, this value has to be taken only as a lower limit. Considering that for the 10x sample the bilayer thickness is $\Lambda=6$ nm we can conclude that, in the textured regions of the multilayer, the crystal registry is maintained at least for few Mg/Ti bilayer repetitions and that the Mg/Ti interfaces are therefore partially coherent, as already suggested by the comparison between experimental and simulated XRD patterns (Fig. \ref{XRDlog}).

\begin{figure}[htbp]
\begin{center}
\includegraphics[width=8.6 cm]{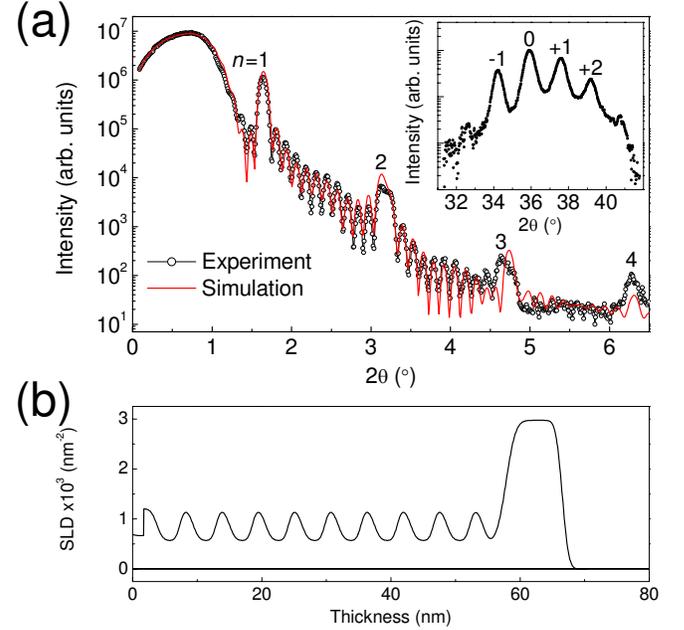}
\caption{(color online) (a) X-Ray Reflectivity pattern of the 10x sample, 10x[Ti(2 nm)Mg(4 nm)]Pd(10 nm), taken at 333 K in vacuum (base pressure $=10^{-7}$ Pa). The red solid line
is a simulation of the multilayer using GenX.\cite{GenX} The inset shows the XRD pattern measured at high angles on a 10x sample deposited in the same run. (b) Real part of the Scattering Length Density (SLD) profile corresponding to the simulation in Fig. \ref{XRR1}a.}\label{XRR1}
\end{center}
\end{figure}

The 10x sample, 10x[Ti(2 nm)Mg(4 nm)], has been explored in more detail by means of \textit{in-situ} X-Ray Reflectivity at different hydrogen pressures. The XRR measurement of the as-deposited state of sample 10x, covered with 10 nm of Pd, is shown in Fig. \ref{XRR1}a: the measurement is conducted in vacuum (base pressure $=10^{-7}$ Pa) at 333 K and the sample is deposited on a Si substrate with (100) orientation.
From the satellite peaks positions in the XRD pattern measured at high angles on a sample deposited in the same run (inset in Fig. \ref{XRR1}a) we obtain $\Lambda=5.63$ nm and $d_{\mathrm{Mg/Ti}}=0.2500$ nm. The XRR curve is simulated with GenX \cite{GenX} using the model in Fig. \ref{cartoonmultilayers} to obtain the individual thicknesses of the constituent layers and the roughnesses at the
interfaces. The software uses a dynamic optical model that incorporates effects arising from refraction, x-ray absorption, multiple
scattering, instrumental resolution and instrumental geometry. In the fit all the layer thicknesses and the interface roughnesses are varied. A layer of 1.7 nm of SiO$_2$ is added to the simulation on top of the Si (100) substrate to take into account the substrate surface oxidation. The results for the as-deposited state of sample 10x are summarized in the third column of Table \ref{tablexrr}. 
\begin{table}[htbp]
\caption{\label{tablexrr} Nominal and measured structural parameters of the 10x sample, 10x[Ti(2 nm)Mg(4 nm)]Pd(10 nm), as obtained from fitting the XRR patterns measured in vacuum (as-deposited) and in 10$^3$ Pa of H$_2$ (fully hydrogenated): bi-layer thickness $\Lambda$, layers thickness $d$
and roughness $\sigma$.}
\begin{tabular}{c|c|c|c}
							& Nominal		& XRR				& XRR \\
							&				& (vacuum)			& $(\mathrm{p_{H_2}}=10^3$ Pa) \\
							& nm				& nm				              &nm\\
\hline
$\Lambda$					& 6				&5.6					&6.9\\
$d_{\mathrm{SiO_2}}$			& --				&1.7					&2.0\\
$d_{\mathrm{Ti}}$				& 2				&2.0					&2.3\\
$d_{\mathrm{Mg}}$				& 4				&3.6					&4.6\\
$d_{\mathrm{Pd}}$				& 10				&8.7					&8.7\\
$\sigma_{\mathrm{Ti/Mg}}$		& 0				&0.74					&1.0\\
$\sigma_{\mathrm{Mg/Ti}}$		& 0				&0.55					&1.2\\
$\sigma_{\mathrm{Mg/Pd}}$		& 0				&1.2					&1.3\\
$\sigma_{\mathrm{Pd/vacuum}}$	& 0				&0.78					&1.1
\end{tabular}
\end{table}
The agreement between the $\Lambda$ values obtained from the satellite peaks position and from the simulation of the XRR pattern is excellent. The layers are flat with interface roughnesses of the order of one unit cell (assuming hcp Mg with $\mathrm{c}=0.521$ nm and hcp Ti with $\mathrm{c}=0.469$ nm). Fig. \ref{XRR1}b shows the real part of the Scattering Length Density (SLD) profile corresponding to the simulation in Fig. \ref{XRR1}a. The SLD, which is given by the mass density profile times the scattering lengths of the elements, gives an idea of the deviations from a perfect square-wave model that we have to introduce in the simulation, in order to accurately reproduce the experimental measurement.

Another method to determine the bilayer thickness $\Lambda$, in the limit of a kinematic approximation, is by looking at the position of the reflectivity peaks at low angles. We have:\cite{agarwal1991}
\begin{equation}\label{Bragg2}
\sin^2\theta_n=\left(\frac{\lambda_{\mathrm{CuK\alpha_1}}}{2\Lambda}\right)^2n^2+2\delta
\end{equation}
where $n$ is the order of the reflectivity peaks, as shown in Fig. \ref{XRR1}a, and $\delta$ is the deviation from unity of the real part of the average refractive index and, in first approximation, can be neglected.\cite{fullertonPRB1992, Palsson2008} From a linear fit of the plot $\sin^2\theta_n/\lambda^2_{\mathrm{CuK\alpha_1}}$ versus $n^2$ we calculate $\Lambda=5.64$ nm, in agreement with the values obtained both from the XRR simulation and from the satellite peak positions in the XRD pattern.

\begin{figure}[htbp]
\begin{center}
\includegraphics[width=8.6 cm]{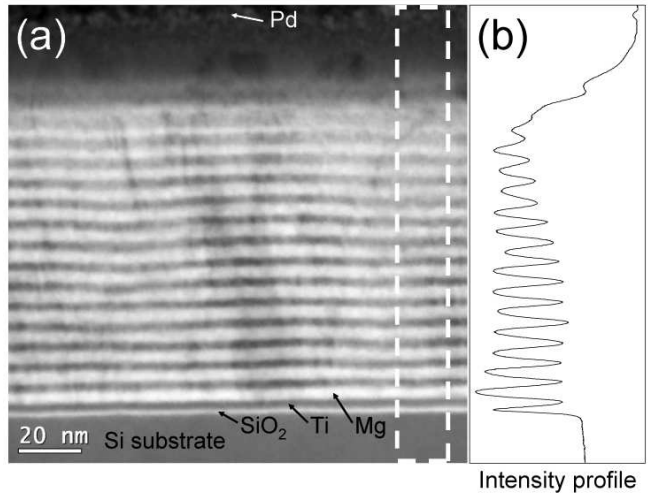}
\caption{(a) Cross-section bright-field STEM image of a 20x[Ti(2 nm)Mg(4 nm)] multilayer deposited on a Si(100) substrate and covered with 10 nm of Pd. (b) Intensity profile of the area delimited by a white dashed line. The maxima correspond to the Mg bright layers, except for the first (bottom) peak, which is attributed to SiO$_2$ at the substrate/film interface.}\label{TEM_asdep1}
\end{center}
\end{figure}

\begin{figure}[htbp]
\begin{center}
\includegraphics[width=8.6 cm]{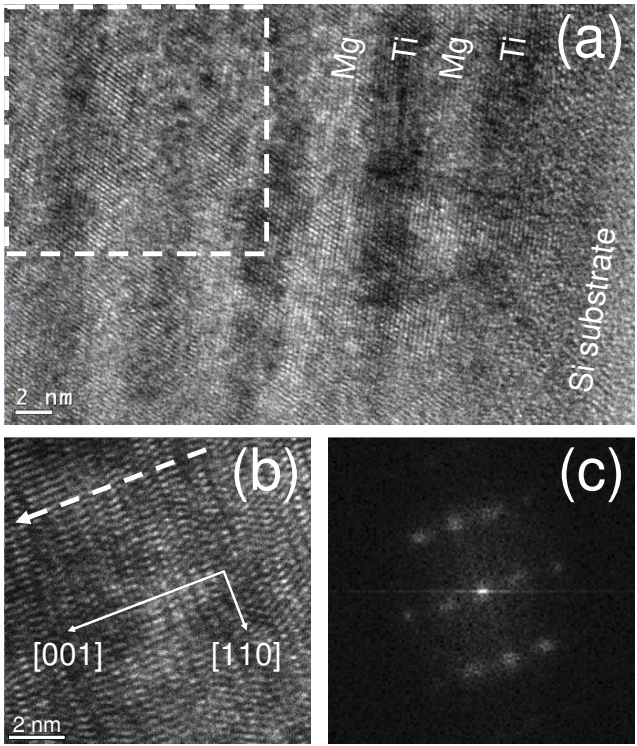}
\caption{(a) Cross-section TEM image of the 20x sample shown in Fig. \ref{TEM_asdep1}. The dashed area highlights a region in which structural coherence is maintained across multiple Mg/Ti interfaces. (b) HRTEM image of sample 20x, taken approximately in the middle of the multilayer. The dashed arrow, parallel to the [001] direction of the hcp structure, indicates the growth direction of the multilayer. (c) Fourier transform of  Fig. \ref{TEM_asdep3}b.}\label{TEM_asdep3}
\end{center}
\end{figure}

In Fig. \ref{TEM_asdep1}a we show the cross-section bright-field Scanning Transmission Electron Microscopy (STEM) image of a 20x[Ti(2 nm)Mg(4 nm)] multilayer, deposited on a Si(100) substrate and covered with 10 nm of Pd.
The sample is well layered with slightly wavy Mg/Ti interfaces. Due to cumulative roughness, only 16 out of 20 bilayer repetitions are visible. Figure \ref{TEM_asdep1}b shows the intensity profile of the area delimited by the white dashed line in Fig. \ref{TEM_asdep1}a. The maxima in the profile correspond to the bright Mg layers, except for the first peak on the bottom, which is attributed to the SiO$_2$ film covering the Si substrate. The effect of cumulative roughness is clearly visible in the intensity profile, where the peak-to-valley ratio decreases with increasing film thickness. Note that the sample measured in TEM is twice as thick as all the other investigated samples, in which the cumulative roughness effect is therefore going to be much smaller. Figure \ref{TEM_asdep3}a shows a cross-section image of the same sample, in which it is clearly visible how the film is partially crystalline, with grains extending for several Mg/Ti repetitions, as already suggested by X-Ray Diffraction results.
Figures \ref{TEM_asdep3}b and \ref{TEM_asdep3}c show a High Resolution TEM (HRTEM) image of the 20x[Ti(2 nm)Mg(4 nm)] multilayer and its Fourier transform, respectively. As expected, the multilayer has hexagonal closed-packed structure and grows along to the [001] direction. Although a slight decrease in crystallinity is observed with increasing thickness (not shown here), the crystal structure and orientation shown in Figure \ref{TEM_asdep3}b are visible across the multilayer from the substrate up to the Pd cap. 

\subsubsection{Hydrogen loading}

When exposed to H$_2$ gas at room temperature, the Pd-capped Mg/Ti multilayers hydrogenate in two consecutive steps: (i) at lower H$_2$
pressures only the Ti layers form a hydride while Mg remains in its metallic state, (ii) at higher H$_2$ pressures also Mg absorbs hydrogen
forming MgH$_2$. Such loading sequence is due to the lower (more negative) enthalpy of hydride formation of TiH$_2$ (-65 kJ(mol H)$^{-1}$)\cite{Manchester} with
respect to MgH$_2$ (-37.2 kJ(mol H)$^{-1})$\cite{stampferJACS1960} and can be detected from XRD, XRR and optical spectroscopy measurements.

\begin{figure}[htbp]
\begin{center}
\includegraphics[width=8.6 cm]{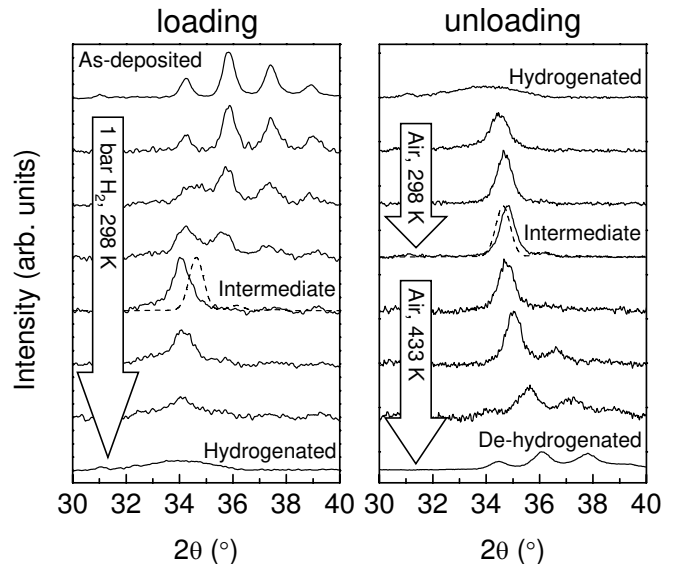}
\caption{XRD patterns measured during loading (left) and unloading (right) of the 10x sample, 10x[Ti(2 nm)Mg(4 nm)]Pd(10 nm). The intensities are normalized by the measuring time. The dashed lines are simulations of a perfect hcp-(002) Mg / fcc-(111) TiH$_2$ 10x multilayer.}\label{loadingunloading}
\end{center}
\end{figure}

Figure \ref{loadingunloading} exhibits the XRD patterns measured during loading and unloading of sample 10x.
Upon loading the sample is exposed to 1 bar H$_2$ pressure at room temperature. The loading sequence shows an intermediate step in which only one peak is visible. This intermediate peak is similar to what is expected for a perfect hcp-(002) Mg / fcc-(111) TiH$_2$ multilayer, (dashed lines in Fig. \ref{loadingunloading}), although it is shifted to slightly lower angles, suggesting that Mg layers might also be partially hydrogenated. The final hydrogenated state shows poor crystallinity with a broad reflection at $2\theta\approx34^{\circ}$. When unloading the multilayer in air at room temperature the intermediate peak is recovered, suggesting that, due to kinetic limitations, hydrogen does not desorb from the Ti layers in the sample. In order to recover the initial metallic state the sample has to be heated in air at 433 K. Although the satellite peaks in the final de-hydrogenated state are broader than in the as-deposited initial sample, it is remarkable that crystallinity appears again, after the large changes in volume upon hydrogenation and dehydrogenation of the Mg and Ti layers.

\begin{figure}[htbp]
\begin{center}
\includegraphics[width=8.6 cm]{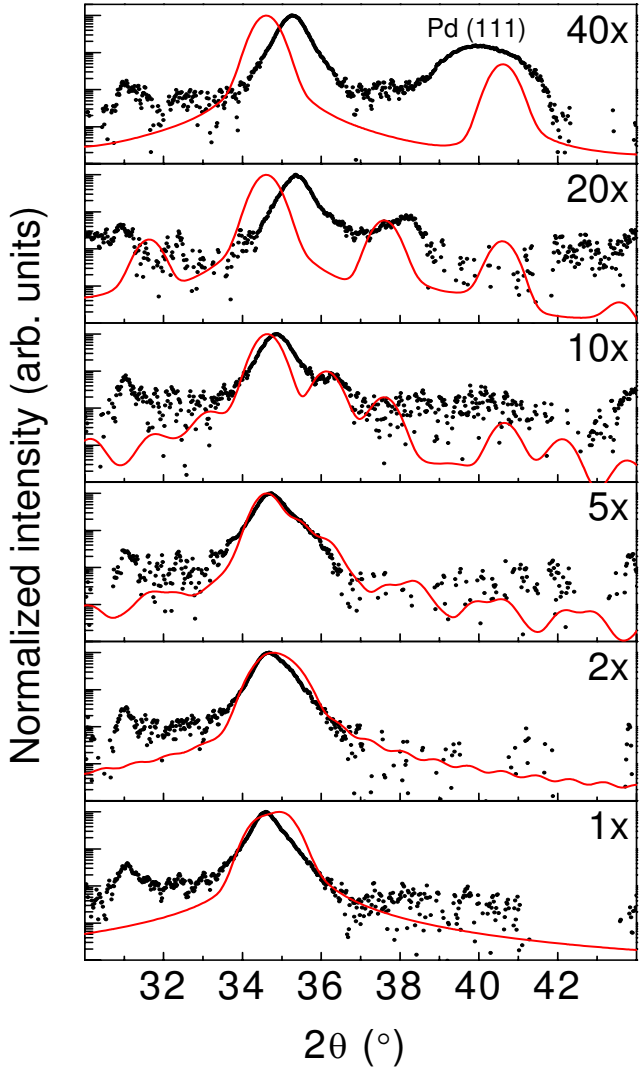}
\caption{(color online) Measured (dots) and simulated (lines) XRD patterns for the partially desorbed, ``intermediate'' states of the samples described in the present work. The simulations assume samples geometries identical to the original multilayers but with fcc-(111) TiH$_2$ instead of hcp-(002) Ti.}\label{alldesorbedlog}
\end{center}
\end{figure}

An intermediate state, similar to the one observed upon desorption in air at room temperature for the 10x sample, corresponding to a hcp-(002) Mg / fcc-(111) TiH$_2$ multilayer, is obtained for the desorbed state of all the multilayers explored in the present work, Fig. \ref{alldesorbedlog}. The lines are simulations based on the  ideal step model, in which we assume that the samples have an ideal geometry and that only the Ti layers are hydrogenated.
The agreement between measured and simulated patterns is good. Significant discrepancies in the peaks positions only occur for the 20x and 40x samples, for which the measured peak lies at higher angles with respect to the simulated one, indicating that partial release of hydrogen has already occurred from the Ti layers. This is not surprising as the 20x and 40x samples have Ti layers as thin as 1 and 0.5 nm, respectively. In the simulations of Fig. \ref{alldesorbedlog} the gaussian distribution fwhm is increased to $0.6^{\circ}$ to account for reduced crystallite size. Significantly, while the fwhm value reproduces well the width of the fundamental peak for the 40x, 20x, 10x and 5x Mg/TiH$_2$ multilayers, it is too high for the 2x and 1x samples. In the latter the coherence length must therefore be larger, thanks to the reduced amount of Mg/Ti interfaces.

\begin{figure}[htbp]
\begin{center}
\includegraphics[width=8.6 cm]{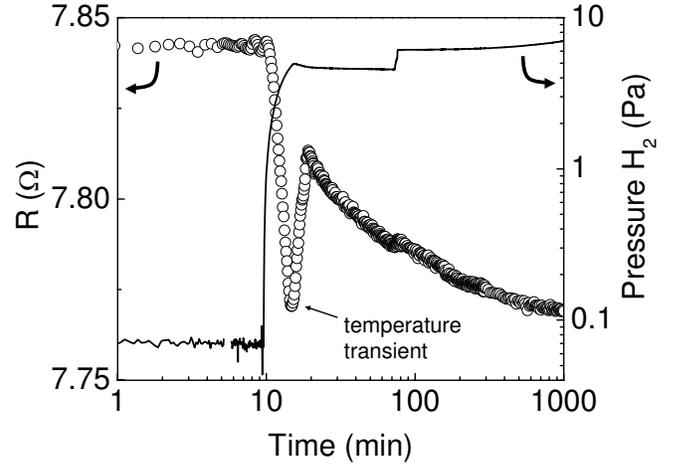}
\caption{Time evolution of the electrical resistance of sample 10x, 10x[Ti(2 nm)Mg(4 nm)]Pd(10 nm), while loading in $\sim$6 Pa of hydrogen at 333 K in the \textit{in-situ} XRR setup.}\label{resistance_first}
\end{center}
\end{figure}

\begin{figure}[htbp]
\begin{center}
\includegraphics[width=8.6 cm]{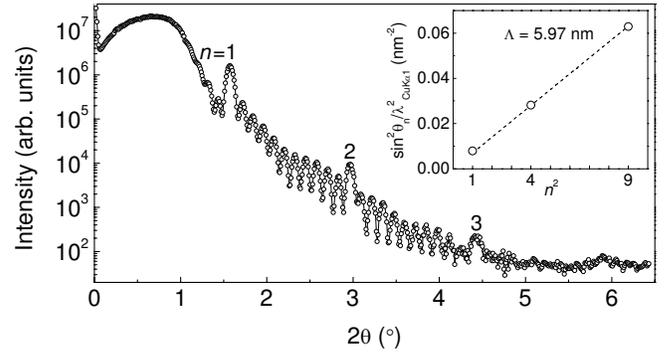}
\caption{XRR from the partially loaded 10x multilayer at $\mathrm{p_{H_2}}=6$ Pa and $\mathrm{T}=333$ K. The inset shows the measured linear dependence of the ratio $\sin^2\theta_n/\lambda^2_{\mathrm{CuK\alpha_1}}$ versus $n^2$.}\label{XRRinter}
\end{center}
\end{figure}

The intermediate state, characterized by the loading of the Ti layers only, has been explored in detail by measuring XRR on a 10x sample exposed to $\sim$6 Pa of hydrogen at 333 K. Such a low pressure is enough to induce hydrogen absorption in Ti but not in Mg. Figure \ref{resistance_first} shows the change in electrical resistance occurring upon loading of the 10x sample in the \textit{in-situ} XRR setup at 333 K under $\sim$6 Pa of hydrogen.
The resistance decreases as a confirmation that only the Ti layers are loaded: unlike MgH$_2$, which is an insulator, TiH$_2$ is a metal with a higher electrical conductivity than Ti.\cite{itoJALCOM2006} The temperature transient highlighted in Fig. \ref{resistance_first} is due to the injection of hydrogen gas at room temperature, in the sample chamber at 333 K. Figure \ref{XRRinter} shows the XRR pattern measured on the 10x sample at 333 K in $\sim$6 Pa of hydrogen. 
From the position of the reflectivity peaks and applying eq. \ref{Bragg2}, we calculate a period expansion of about 5.9$\%{}$, with respect to the as-deposited metallic state (see inset in Fig. \ref{XRRinter}). According to the literature values for the molar volumes of Ti ($\overline{\mathrm{V}}_{\mathrm{Ti}}=10.64$ cm$^3$/mol) and TiH$_2$ ($\overline{\mathrm{V}}_{\mathrm{TiH_2}}=13.3$
cm$^3$/mol), the hydrogenation of titanium should expand the lattice of the host metal by 25\%{}. Given the 3.6:2 thickness ratio of Mg and Ti measured by XRR (Table \ref{tablexrr}) on the as-deposited sample, a uniaxial vertical expansion of the period $\Lambda$ of 5.9\%{}, due to the hydrogenation of Ti only, implies a vertical expansion of the Ti layers of 16.5\%{}. This can only be understood taking into account strong plastic deformations and out-of-plane material pile up, due to the hydrogen-induced in-plane coherency stress in the titanium layers.\cite{laudahnJALCOM1999} This dramatic material movement is likely to remove the structural coherence at the Mg/Ti interfaces and to be at the origin of the ``scissor'' effect observed in Mg layers sandwiched between Ti thin films.\cite{baldiPRL2009,baldiAPL2009}

\begin{figure}[htbp]
\begin{center}
\includegraphics[width=8.6 cm]{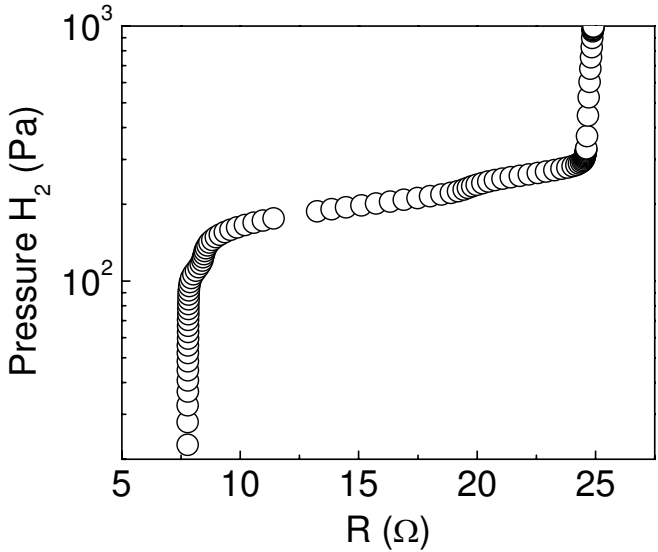}
\caption{Pressure-resistance-isotherm measured during loading of the 10x sample, 10x[Ti(2 nm)Mg(4 nm)]Pd(10 nm), at 333 K in the \textit{in-situ} XRR setup.}\label{resistance}
\end{center}
\end{figure}

\begin{figure}[htbp]
\begin{center}
\includegraphics[width=8.6 cm]{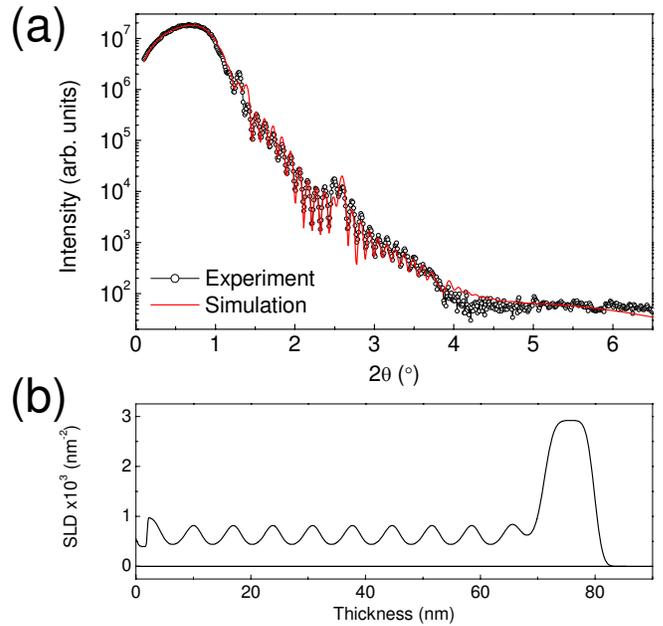}
\caption{(color online) XRR pattern of the fully loaded 10x sample at $\mathrm{p_{H_2}}=10^3$ Pa and $\mathrm{T}=333$ K. The
solid red line is a simulation of the multilayer using GenX.\cite{GenX} (b) Scattering Length Density (SLD) profile corresponding to the simulation in Fig. \ref{XRRhydr}a.}\label{XRRhydr}
\end{center}
\end{figure}

\begin{figure}[htbp]
\begin{center}
\includegraphics[width=8.6 cm]{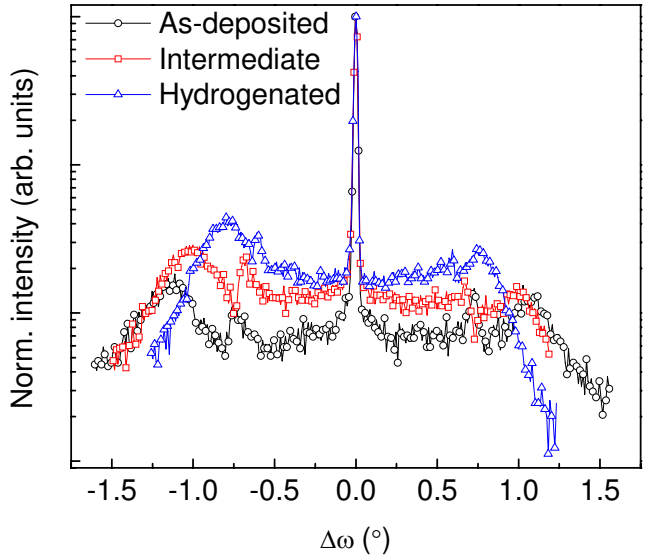}
\caption{(color online) Rocking curves measured over the second order reflectivity peak, for the as-deposited, intermediate and fully hydrogenated 10x sample. The scans are offset and only 50\%{} of the experimental points are shown for clarity.}\label{rocking_XRR}
\end{center}
\end{figure}

After measuring the XRR of the intermediate state in $\sim$6 Pa of H$_2$ at 333 K, we slowly increased the hydrogen pressure up to 10$^3$ Pa at constant temperature, while measuring the electrical resistance of the film (see Fig. \ref{resistance}).
The abrupt increase of electrical resistance between 100 and 300 Pa is due to hydrogen absorption in the Mg layers. After reaching equilibrium in 10$^3$ Pa of H$_2$ we measured the XRR pattern for the fully hydrogenated state, as shown in Fig. \ref{XRRhydr}.
The fit parameters used in the simulation in Fig. \ref{XRRhydr}a are given in the fourth column of Table \ref{tablexrr}. Upon hydrogenation the bilayer thickness $\Lambda$ increases by 23\%{}, from 5.6 nm in the as-deposited state to 6.9 nm in the fully hydrogenated state. Titanium layers expand by 15\%{}, from 2.0 to 2.3 nm, in good agreement with the value obtained from the period expansion observed upon hydrogenation at 6 Pa (16.5\%{}). Magnesium layers expand by $\sim$28\%{}, from 3.6 to 4.6 nm. In the hydrogenation of bulk Mg the molar volume increases by 30\%{}, going from 13.97 cm$^3$/mol in Mg to 18.2 cm$^3$/mol in MgH$_2$: a vertical expansion of the Mg layers of 28\%{} is therefore an indication of strong plastic deformations. Fig. \ref{rocking_XRR} shows the rocking curves measured over the second order reflectivity peak, for the as-deposited, intermediate and fully hydrogenated sample.
The pattern consists of two components, one narrow with a fwhm of 0.023$^{\circ}$ and one much broader. The narrow peak is the specular reflectivity and the width is what would be expected from an optically flat sample. The broader component comes from off-specular scattering which includes scattering from roughness at the interfaces. As can be seen in the figure, the off-specular contribution increases during loading, indicating an increase in interfacial roughness.\cite{savageJAP1991,zabelAPA1994} This is consistent with the reflectivity simulations which also showed a roughness increase upon loading. This roughness increase could in principle also be attributed to atomic interdiffusion, however, since interdiffusion causes a lateral roughness with no particular length scale, it contributes zero or a constant amount to the off specular scattering. The increase in the off-specular scattering observed in Fig. \ref{rocking_XRR} is therefore due to an increase in roughness in the form of thickness variations. The small peaks at either side of the specular reflections are likely to be due to off-specular scattering from the first order reflectivity peaks and not to correlations in the roughness.

\begin{figure}[htbp]
\begin{center}
\includegraphics[width=8.6 cm]{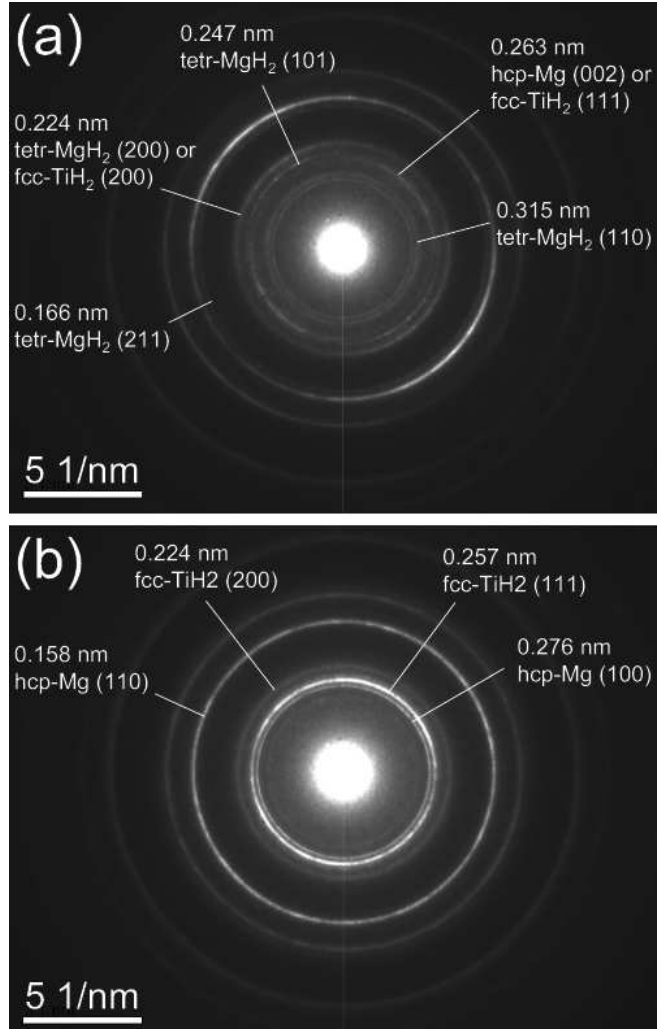}
\caption{Selected area electron diffraction patterns for the hydrogenated state of a 10x Mg/Ti multilayer. (a) Short and (b) long beam exposure.}\label{TEM}
\end{center}
\end{figure}

Due to the poor crystallinity of the fully loaded samples a detailed structural characterization of the hydrogenated state cannot be achieved with XRD only. In order to measure the crystal phases present in the hydrogenated multilayers we performed selected area electron-diffraction patterns of a 10x sample covered with only 1 nm of Pd. Such small amount of Pd allows loading of the sample but its catalytic activity is rapidly suppressed by strong metal-support interaction (SMSI) effects\cite{BorgschultePRB} and it is not enough to promote significant hydrogen desorption.\cite{BorsaPRB2007} The measurements were performed both for few seconds and for 10 minutes of electron beam exposure. In the former case both tetragonal MgH$_2$ and face-centered-cubic TiH$_2$ peaks are observed, while in the latter hydrogen desorption from the MgH$_2$ layers occurs due to electron irradiation and reflections from hexagonal-closed-packed Mg are visible, Fig. \ref{TEM}.
It is noteworthy that in the hydrogenated state no signs of cubic MgH$_2$ are present: in codeposited Mg$_y$Ti$_{1-y}$ thin films, the hydrogenated state is tetragonal for $y>0.87$ and face-centered-cubic for $y<0.87$.\cite{BorsaPRB2007} This fcc phase is similar to the one proposed for high pressure $\beta$-MgH$_2$ \cite{Vajeeston2002} and originates from the structurally coherent dispersion of Mg-rich and Ti-rich nanosized domains in Mg-Ti thin film alloys.\cite{gremaudPRB2008,baldiIJHE2009} The multilayers studied in the present work contain 40 at.\%{} of Ti and one would therefore expect a similar cubic structure to occur upon hydrogenation. Apparently, however, the cubic hydride phase can only be stabilized by a very fine 3D dispersion of Mg and Ti atoms. A 10x multilayer, consisting of repetitions of 2 nm of Ti and 4 nm of Mg, is already too ``segregated'' and leads to standard tetragonal MgH$_2$. This result is consistent with the ``scissor'' effect observed in Ti-sandwiched Mg film, according to which no elastic interaction exists between adjacent Mg and Ti layers.\cite{baldiAPL2009}

\subsection{Optical spectroscopy}

\begin{figure}[htbp]
\begin{center}
\includegraphics[width=8.6 cm]{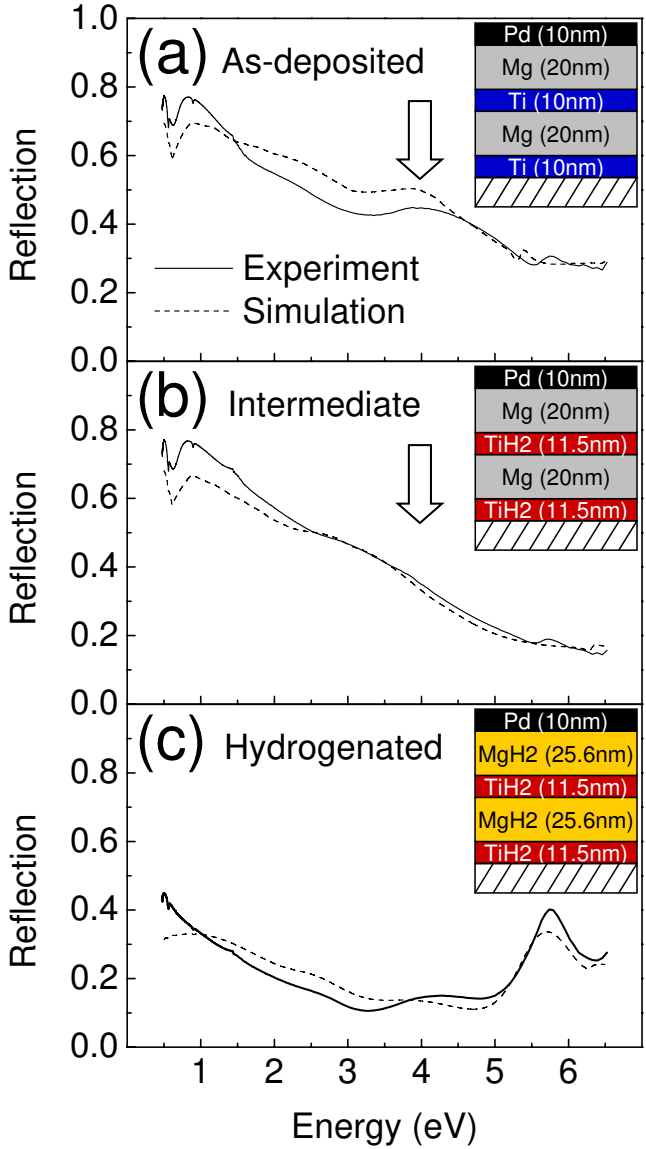}
\caption{(color online) Experimental (lines) and simulated (dots) reflection spectra measured through the quartz substrate for the 2x sample, 2x[Ti(10 nm)Mg(20 nm)]Pd(10 nm), exposed to slowly increasing pressures of a 6\%{}H$_2$/Ar mixtures at room temperature. (a) As-deposited, (b) intermediate and (c) hydrogenated state. The
dielectric functions of Ti, Mg, TiH$_2$, MgH$_2$ and Pd used in the simulations are the same used in Ref. 14.}\label{loadingPE_2x}
\end{center}
\end{figure}

The loading sequence is further investigated by optical spectroscopy measurements. An example is given for sample 2x in Fig.
\ref{loadingPE_2x}, where the measured optical reflection of the multilayer, deposited on a quartz substrate and covered with 10 nm of Pd, is shown for different stages of hydrogenation.
Comparison with simulated optical spectra obtained with SCOUT \cite{TheissSCOUT} shows that the disappearance of the reflection hump at
$\sim$4 eV at the beginning of the hydrogenation process is due to the formation of TiH$_2$ (see white arrow in Fig. \ref{loadingPE_2x}a
and \ref{loadingPE_2x}b). In the simulations shown in Fig. \ref{loadingPE_2x} the thickness increase of the individual Ti and Mg layers has
been taken according to the variations observed by means of XRR, $\sim$15\%{} for Ti and 28\%{} for Mg, leading to
an excellent agreement with the measured spectra.
\vspace{24pt}

\subsection{Hydrogenography}

\begin{figure}[htbp]
\begin{center}
\includegraphics[width=8.6 cm]{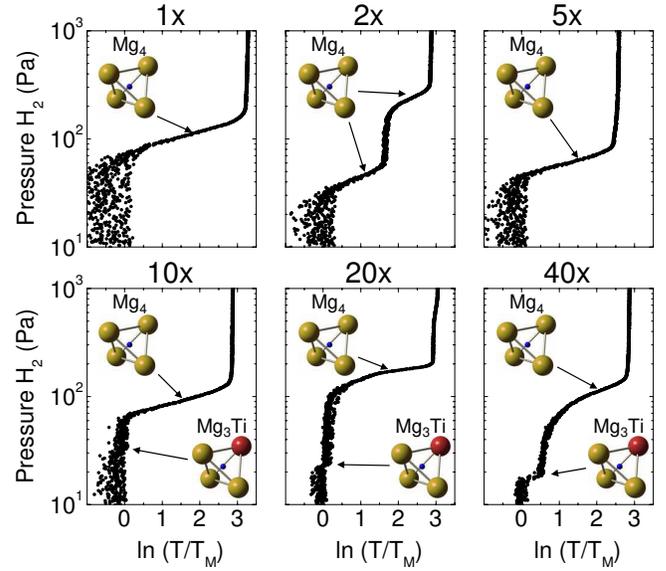}
\caption{(color online) Loading Pressure-Transmission-Isotherms measured by hydrogenography at 333 K for the Mg/Ti multilayers covered with 10 nm of Pd.}\label{h2ography}
\end{center}
\end{figure}

In order to interpret the Pressure-optical Transmission-Isotherms (PTIs) measured by hydrogenography on the Pd-capped Mg/Ti multilayers we make use of the following assumptions: 1) Mg layers in direct contact with Pd feel an elastic constraint, due to the formation of Mg-Pd alloys at the interface, which leads to plateau pressures higher than what expected from bulk Mg ($p^{333\ \mathrm{K}}_{\mathrm{bulk\  Mg}}=12$ Pa\cite{Krozer1990});\cite{baldiPRL2009} 2) in the process of Mg-Pd alloy formation typically 6 nm of Mg are ``lost'' and do not contribute to the optical change occurring upon hydrogen absorption;\cite{baldiPRL2009} 3) Mg films surrounded by Ti layers, on the contrary, do not feel significant elastic constraints, thanks to the positive enthalpy of mixing of Mg and Ti which leads to poor interface adhesion. Furthermore Ti absorbs hydrogen at lower pressures than Mg and the consequent lattice expansion removes the partial lattice coherence at the Ti/Mg interfaces and leads to quasifree Mg layers.\cite{baldiAPL2009} The PTIs measured at 333 K for the multilayers studied in the present work are shown in Fig. \ref{h2ography}.
The PTI for the 1x sample, Ti(20 nm)Mg(40 nm)Pd(10 nm), shows a relatively high plateau pressure with respect to bulk Mg ($p^{333\ \mathrm{K}}_{\mathrm{bulk\  Mg}}=12$ Pa),\cite{Krozer1990} due to the elastic clamping of the top Pd layer.\cite{baldiPRL2009} The same effect, together with the ``scissor'' property of Ti,\cite{baldiPRL2009} is responsible for the double plateau observed for the 2x sample, Ti(10 nm)Mg(20 nm)Ti(10 nm)Mg(20 nm)Pd(10 nm): the bottom Mg layer (lower plateau) is sandwiched between two Ti layers and it is therefore quasifree,\cite{baldiAPL2009} while the top one (higher plateau) is in contact with the Pd cover and feels its elastic constraint, therefore loading at higher hydrogen pressures. Following the same line of reasoning one would expect for the 5x sample to find two plateaus: a plateau at low pressure, build up by the bottom 4 Mg layers sandwiched between Ti and a plateau at higher pressure, coming from the top Mg layer in contact with Pd. The top plateau however is not visible due to the fact that the uppermost Mg layer is only 5 nm thick and therefore completely alloyed to Pd. In the 10x 20x and 40x the plateau pressure slightly increases with respect to the 5x sample. A possible explanation for such behavior is that in the 10x, 20x and 40x samples the individual Mg layers are very thin (4, 2 and 1 nm, respectively) and surface energy might provide a relevant contribution to the enthalpy of hydride formation.\cite{berubeIJHE2008} Another effect that might play a role is the fact that in these samples the Ti layers (2, 1 and 0.5 nm) might not be thick enough to form perfectly closed films: the Mg layers would therefore not be completely shielded from the clamping effect of the top Pd cover. For the 40x sample a sloping plateau similar to the one measured for a co-deposited Mg$_{0.6}$Ti$_{0.4}$ alloy is observed,\cite{Gremaud2007ADM} suggesting that the microstructure of this sample is closer to a mixed alloy than to a well defined multilayer. 

It is noteworthy that for the 10x, 20x and 40x samples a small plateau appears at very low hydrogen pressures. The width of this plateau is proportional to the number of Mg/Ti interfaces present in the multilayers. We interpret these plateaus as originating from hydrogen atoms populating the interstitial sites located at the Mg/Ti interfaces: in the metallic films the hydrogen atoms are located in interstitial tetrahedral sites, inside the Mg layers these are Mg$_4$ sites but, crossing the Mg/Ti interfaces, there will also be Mg$_3$Ti, Mg$_2$Ti$_2$, MgTi$_3$ and Ti$_4$ sites. Since the formation enthalpy of TiH$_2$ is smaller than the one of MgH$_2$, the substitution of one (or more) Mg atoms with Ti in a tetrahedral site will lower the absorption energy and the site will be populated at lower hydrogen pressures.\cite{gremaudPRB2008} As can be seen in Fig. \ref{resistance}, a similar small plateau at the beginning of the hydrogenation process is also observed when measuring the film electrical resistance, while loading the 10x sample used in the XRR measurement. A comparison of Fig. \ref{resistance} and Fig. \ref{h2ography} shows that, when loaded in the \textit{in-situ} XRR setup, the 10x sample absorbs hydrogen at a higher pressure with respect to the loading in the hydrogenography optical setup. This is not surprising since, when measuring isotherms by hydrogenography, we want to obtain equilibrium curves and the pressure is increased very slowly in 20 hours. In the XRR setup on the other hand the loading was completed in 30 minutes and it is therefore not an equilibrium measurement. The small step observed at $\sim$230 Pa in the pressure-resistance-isotherm cannot be explained with simple thermodynamic considerations, but it could be related to the fact that the measurement is done in the kinetic regime, as such a step is not observed in the equilibrium isotherm obtained by hydrogenography.

\subsection{Diffusion simulation of hydrogen cycling}

While loading the multilayers in the XRD setup we exposed them to 1 bar of H$_2$ at room temperature, a pressure sufficient to hydrogenate both the Ti and the Mg layers. Nevertheless we observed an intermediate state in which only the Ti layers were loaded. Furthermore, upon exposure to air at room temperature hydrogen desorbs from the Mg layers but remains trapped in the Ti ones and the original metallic state can only be recovered by heating the samples up to 433 K. These loading and unloading sequences can be qualitatively explained by looking at the difference in thermodynamic properties of hydrogen absorption in Mg and Ti. Pasturel \textit{et al.}\cite{pasturelCM2007} have already shown how the chemical potential of hydrogen in different transition metals (TM) can influence the hydrogen sorption kinetics in Mg$_2$Ni/TM/Pd trilayers. Here we extend their model to a system with 7 layers, in order to account for the multiple repetitions typical of a multilayer. The model treats hydrogen dissolved in metals in the lattice-gas approximation, without including any H-H interaction. The chemical potential of hydrogen in a metal ($M$) is therefore written as:
\begin{eqnarray}\label{chempot}
\mu^{\mathrm{M}}_{\mathrm{H}}(c_{\mathrm{M}})&=&\overline H_{\mathrm{H}}-T\overline S_{\mathrm{H}}\nonumber\\
&=&\overline H_{\mathrm{H}}-T\left(\overline S_{\mathrm{conf.}}+\overline S_{\mathrm{vibr.}}\right)
\end{eqnarray}
$\overline H_{\mathrm{H}}$ and $\overline S_{\mathrm{H}}$ are the partial molar enthalpy and entropy of hydrogen in metals, respectively. The configurational contribution to the entropy, $\overline S_{\mathrm{conf.}}$, is proportional to $\ln\left[c_{\mathrm{M}}/\left(1-c_{\mathrm{M}}\right)\right]$,  where $c_{\mathrm{M}}$ is the hydrogen concentration in the metal. This term takes care that hydrogen atoms obey Fermi-Dirac statistics in the host metallic lattice, due to the single occupation of interstitial sites. In most metals the vibrational term, $\overline S_{\mathrm{vibr.}}$, is small at moderate temperatures and can be neglected, with the significant exception of Pd ($\overline S_{\mathrm{vibr.}}\approx19.2$ JK$^{-1}$(mol H)$^{-1}$ at 298 K).\cite{gremaudphdthesis}

Each iteration of the model consists of two steps. First, the concentration within each layer is updated according to the diffusion equation.  Subsequently, we impose the equality of chemical potential at the interface between two materials by changing the concentrations at the sites adjacent to the interface such that the chemical potential is made equal on both sites, while keeping the \textit{total} amount of hydrogen atoms unchanged. Note that the equality in chemical potentials at the interfaces does not imply the equality of hydrogen concentrations in the adjacent surfaces of the two layers. Instead of using the experimental values of the hydrogen pressure and the enthalpies and entropies of hydrogen absorption in Mg, Ti and Pd we define the following dimensionless parameters:
\begin{eqnarray}\label{model}
e_{\mathrm{H}}&=&\frac{1}{2}\ln\frac{p_{\mathrm{H_2}}}{p^0}\nonumber\\
e_{\mathrm{M}}&=&\frac{\mu^{\mathrm{M}}_{\mathrm{H}}-\frac{1}{2}\mu^0_{\mathrm{H_2}}}{RT}\nonumber\\
&=&\frac{\overline{H}_{\mathrm{H}}-T\overline{S}_{\mathrm{H}}-\frac{1}{2}H^0_{\mathrm{H_2}}+\frac{1}{2}TS^0_{\mathrm{H_2}}}{RT}\nonumber\\
&=&\frac{\left(\overline{H}_{\mathrm{H}}-\frac{1}{2}H^0_{\mathrm{H_2}}\right)-T\left(\overline{S}_{\mathrm{H}}-\frac{1}{2}S^0_{\mathrm{H_2}}\right)}{RT}\nonumber\\
&=&\frac{\Delta\mathrm{H}_{\mathrm{M}}}{RT}-\frac{\Delta S^0_{\mathrm{M}}}{R}
\end{eqnarray}
where $p^0=10^5$ Pa, $\Delta S^0_{\mathrm{M}}$ is the entropy of hydride formation which, except for Pd ($\Delta S^0_{\mathrm{Pd}}=-48.7$ JK$^{-1}$(mol H)$^{-1}$), is taken equal to the entropy of hydrogen gas at standard pressure ($\Delta S^0_{\mathrm{M}}\approx-\frac{1}{2}S^0_{\mathrm{H_2}}=-65$ JK$^{-1}$(mol H)$^{-1}$) and $\Delta H_{\mathrm{M}}$ is the enthalpy of hydride formation: $\Delta\mathrm{H}_{\mathrm{Pd}}=-20.5$ kJ(mol H)$^{-1}$,\cite{Frieske1973} $\Delta\mathrm{H}_{\mathrm{Mg}}=-37.2$ kJ(mol H)$^{-1}$,\cite{stampferJACS1960} $\Delta\mathrm{H}_{\mathrm{Ti}}=-65$ kJ(mol H)$^{-1}$.\cite{Manchester} Substituting these values in eq. \ref{model} and taking T = 333 K, we obtain: $e_{\mathrm{Pd}}=-1.5$, $e_{\mathrm{Mg}}=-5.6$ and $e_{\mathrm{Ti}}=-15.6$.

\begin{figure}[htbp]
\begin{center}
\includegraphics[width=8.6 cm]{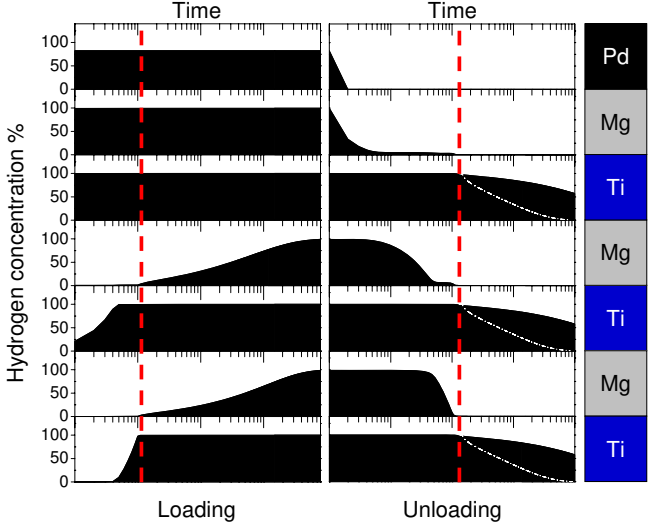}
\caption{(color online) Simulated time evolution (in logarithmic scale) of the hydrogen concentration in each layer of a 3x[Ti($z$)Mg(2$z$)]Pd($z$) sample, upon hydrogen absorption ($e_{\mathrm{H}}=0$) and desorption ($e_{\mathrm{H}}=-20$). The vertical dashed lines represent the intermediate states, observed experimentally both upon loading and unloading, in which only the Ti layers are hydrogenated. The white dash-dot lines are the simulated desorption rate from the Ti layers at 433 K. }\label{diffusion}
\end{center}
\end{figure}

In Fig. \ref{diffusion}, we simulate the loading and unloading behavior of a multilayer made of 3 Ti/Mg repetitions and covered with a Pd layer, in which the Mg layers are twice as thick as the Pd and Ti ones: 3x[Ti($z$)Mg(2$z$)]Pd($z$). In order to highlight only the effect of the chemical potential, we assume that the hydrogen diffusion coefficient is the same for all the materials. The loading pressure is taken as 10$^5$ Pa ($e_{\mathrm{H}}=0$) and the unloading pressure is taken negative enough in order to allow desorption from all the layers ($e_{\mathrm{H}}=-20$).
Upon exposure to 10$^5$ Pa of H$_2$ the Pd and Ti layers and the uppermost Mg layer load very quickly, while the Mg layers ``sandwiched'' between Ti absorb hydrogen more slowly, effectively producing a transient intermediate TiH$_2$/Mg multilayer. This is due to the fact that at the beginning of the loading process, when the hydrogen concentration is low everywhere, Ti acts as a hydrogen sink due to its lower enthalpy of hydride formation, effectively sucking hydrogen atoms lying at the Ti/Mg interfaces. When the Ti layers are almost full the logarithmic entropic term in eq. \ref{chempot} dominates over the energetic term and Mg layers start to absorb hydrogen. Upon unloading the effect is opposite: at the beginning the hydrogen concentration is very high everywhere and hydrogen atoms remain trapped in the Ti layers until the adjacent Mg layers are ``empty'' enough to equilibrate the chemical potentials. When desorption starts in the Ti layers, however, the hydrogen concentration in the Mg layers is so low that the total flux of hydrogen atoms through Mg (which is proportional to the gradient in concentration) is minimal, resulting in a very slow hydrogen desorption. In Fig. \ref{diffusion} we also show that increasing the temperature of the system to 433 K enhances the rate of hydrogen desorption from the Ti layers (white dash-dot lines). With these simple simulations we can therefore give a qualitative interpretation of the persistence of hydrogen in the Ti layers, both upon loading and unloading, as observed experimentally.

\section{Conclusions}
We have prepared several Pd-capped Mg/Ti multilayers with various periodicities by means of magnetron sputtering. The deposited samples have lattice parameters in the $z$ direction close to their nominal bulk values and low interfacial roughnesses. Partial structural coherence exists at the Mg/Ti interfaces in the as-deposited state but it is reduced upon hydrogen absorption and desorption. The hydrogen loading sequence, as confirmed by XRD, XRR and optical spectroscopy, agrees with what is expected from thermodynamic considerations on the enthalpies of formation of magnesium and titanium hydrides: Mg/Ti $\rightarrow$ Mg/TiH$_2$ $\rightarrow$ MgH$_2$/TiH$_2$. Hydrogen absorption in both the Ti and Mg layers leads to large expansions in the vertical out-of-plane direction, well beyond the elastic regime, indicating that massive material pile up, due to plastic deformations and creation of defects, has to take place. Upon dehydrogenation hydrogen is kinetically trapped in the Ti layers. Complete desorption only occurs upon exposure to air at 433 K leading to the original metallic layered structure, with a shorter coherence length. Magnesium hydride in a Mg/Ti multilayer with period $\Lambda$ as small as 6 nm, retains its standard tetragonal structure, suggesting that the occurrence of a cubic hydrogenated phase, as observed in partially segregated Mg$_y$Ti$_{1-y}$ ($y<0.87$) thin films,\cite{gremaudPRB2008,baldiIJHE2009} can only be stabilized by a very fine dispersion of Mg-rich and Ti-rich domains. Pressure-optical Transmission-Isotherms measured by hydrogenography, can be interpreted on the basis of the clamping effect on thin Mg films due to the adjacent layers\cite{baldiPRL2009,baldiAPL2009} and, possibly, on the surface energy differences between the metallic and hydrogenated states of ultra thin Mg films. A simple diffusion model allows us to reproduce both the loading and unloading sequences measured experimentally. 

\section{Acknowledgments}
This work is financially supported by the Technologiestichting STW, the Nederlandse Organisatie voor Wetenschappelijk Onderzoek (NWO) through the Sustainable Hydrogen Programme of Advanced Chemical Technologies for Sustainability (ACTS) and the Marie Curie Actions through the project COSY:RTN035366. One of the authors (B.H.) acknowledges support from Knut and Alice Wallenberg foundation. We thank S. De Man, V. Palmisano and L. Mooij for fruitful discussion.

\end{document}